\documentclass[twocolumn,showpacs,preprintnumbers,showkeys,superscriptaddress]{revtex4-1}
\usepackage{CJK}
\usepackage{mathrsfs}
\usepackage{amssymb}
\usepackage{amsmath}
\usepackage{graphicx}
\usepackage{dcolumn}
\usepackage{bm}
\usepackage{graphicx}
\usepackage{float}
\usepackage[normalem]{ulem}
\usepackage{color}
\usepackage{multirow}

\begin{document}
\begin{CJK*}{UTF8}{}
\title{Phase change near $N=70$ in the wave function of the $I^{\pi}=11/2^-$ isomers along the cadmium-isotope chain}

\author{Y. Lei ({\CJKfamily{gbsn}雷杨}) }
\email{leiyang19850228@gmail.com}
\affiliation{Key laboratory of neutron physics, Institute of Nuclear Physics and Chemistry, China Academy of Engineering Physics, Mianyang 621900, China}

\author{H. Jiang ({\CJKfamily{gbsn}姜慧})}
\affiliation{School of Arts and Sciences, Shanghai Maritime University, Shanghai 201306, China}
\affiliation{Department of Physics, Shanghai Jiao Tong University, Shanghai 200240, China}

\author{S. Pittel}
\affiliation{{Bartol Research Institute and Department of Physics and Astronomy, University of Delaware, Newark, Delaware 19716, USA }}
\date{\today}

\begin{abstract}
The electromagnetic features of the $11/2^-$ isomers in the $^{111-127}$Cd isotopes are reproduced by numerically optimized shell-model wave-functions. A sudden phase change of the wave functions at $N = 70$ is identified and further confirmed by the evolution of B(E2, $7/2^-_1\rightarrow 11/2^-_1$) values. This phase change gives rise to different linear relations for the $Q$ and $\mu$ values with $N<70$ and $N>70$, as needed to reproduce the experimental data. The particle-hole transformation properties for $h_{11/2}$ neutrons in a well-isolated subshell involving degenerate $s_{1/2}$, $d_{3/2}$, $d_{5/2}$ and $h_{11/2}$ orbits is suggested as a possible explanation for this phase change.
\end{abstract}
\pacs{21.10.Ky, 21.60.Cs, 23.20.-g, 23.20.Js}
\maketitle
\end{CJK*}

\section{introduction}\label{int}
\begin{figure}
\includegraphics[angle=0,width=0.48\textwidth]{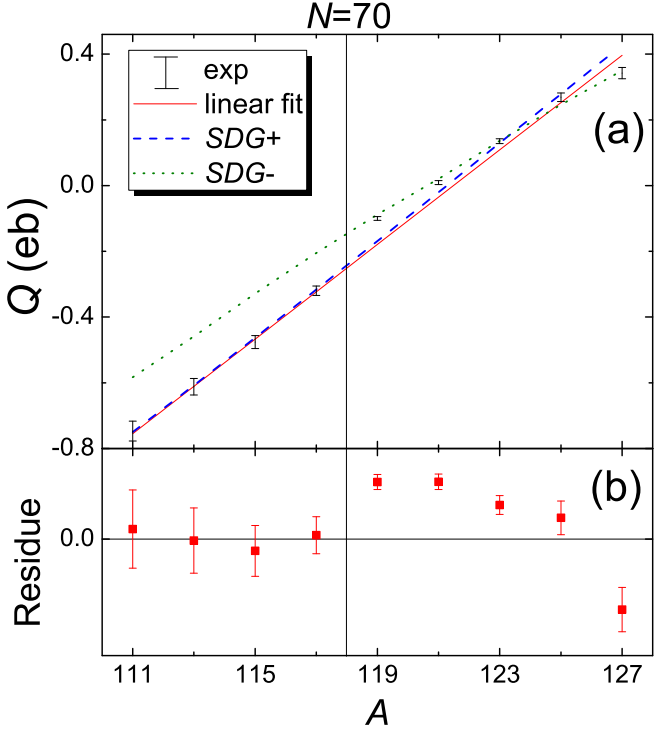}
\caption{(Color online) $Q$ values of the $11/2^-$ isomers in the Cd isotopes. In panel (a), we compare experimental values \cite{cd11}, linear fitting results, and results from the $SDG+$ and $SDG-$ calculations described in the text. Panel (b) presents residues of the linear fit. The $N=70$ position is highlighted. For $N<70$, the residues are smaller than experimental errors, and thus a linear relation follows. On the contrary, the residues for $N>70$ are larger than experimental errors, which means that $Q$ linearity is not respected. The $SDG+$ and $SDG-$ results are consistent with the $N<70$ and $N>70$ experimental data, respectively, as discussed in the text.}\label{q}
\end{figure}

Cadmium ($Z = 48$) is the first isotope chain in the $N = 50$ to $N = 82$ region for which systematic experimental information about excited states is available from one shell closure to the other \cite{cd-first-1,cd-first-2}. The spectral and electromagnetic evolution revealed in these isotopes permits a systematic study of their nuclear structure properties, which also have an impact on nuclear astrophysics \cite{quench-1,quench-2,pro-1,pro-2,oct,astro-1,astro-2}. A recent study of the $I^{\pi}=11/2^-$ isomers in Cd isotopes \cite{cd11} indicates that the quadrupole and magnetic moments (denoted by $Q$ and $\mu$, respectively) both vary roughly as linear functions of neutron number. However, as was also noted, there is a noticeable change in the linear relation for the $\mu$ values at $N=70$.  Indeed, if one performs a linear fit to the $Q$ values, as shown in fig. \ref{q}(a), a similar behavior is observed: the best-fit linear relation perfectly describes $Q$ values with $N<70$, while residues from the same fit obviously exceed experimental errors for $N>70$ as shown in fig. \ref{q}(b). Thus, the $Q$ values also exhibit a change in linear behavior before and after $N=70$.

Recently a density functional calculation \cite{dif-70} suggested that the $Q$ linearity in the Cd isotopes has a different mechanism for $N<70$ and $N>70$. In this work, we study the mechanism for this change in linear behavior of both the $Q$ and $\mu$ values at $N=70$ in the framework of a simple shell-model treatment, to see whether new insight emerges.

\section{calculations}\label{cal}
We begin by outlining the basic ingredients of our shell-model calculations:
\begin{itemize}
\item
[1] We first assume simple, albeit reasonable, proton and neutron configurations from which to build the $11/2^-$ isomers.
\item
[2] We next calculate the $Q$ and $\mu$ matrix elements for the above shell-model configurations.
\item
[3] We consider several trial wave-functions for the isomers and fit their respective amplitudes to optimally describe the $Q$ and $\mu$ values for the $11/2^-$ isomers.
\end{itemize}
As we will see, the procedure leads to a set of wave functions that describes very well the evolution of the quadrupole and magnetic moments of the $11/2^-$ isomers and furthermore suggests a fairly simple picture for the change in behavior at $N=70$.

For the proton configuration, we assume that the proton polarization of the $11/2^-$ isomer derives from the excitation of two proton holes in the $Z = 50$ core, as suggested by ref. \cite{cd11}, and thus include in our proton configuration space the $(g_{9/2})^{-2}$ hole states beyond the $pf$ shell closure. There are five $(g_{9/2})^{-2}$ hole states with total spins $J=0, 2, 4, 6, 8$, which are denoted by $S_{\pi}= (g_{9/2}\otimes g_{9/2})^0$, $D_{\pi}= (g_{9/2}\otimes g_{9/2})^2$, $G_{\pi}= (g_{9/2}\otimes g_{9/2})^4$, $I_{\pi}= (g_{9/2}\otimes g_{9/2})^6$, and $K_{\pi}= (g_{9/2}\otimes g_{9/2})^8$, respectively.

We note that the Cd isotopes are nearly spherical, so that a minimal amount of proton-neutron configuration mixing should be present \cite{pn-def-pittel3}. Since the low-lying excitation has been attributed to protons in ref. \cite{cd11} and as we assume in this work, we neglect the neutron excitation for theoretical self-consistency. As a result, our calculation freezes the neutron configuration to a simple seniority state with one unpaired $h_{11/2}$ neutron (denoted by $\phi_\nu$) as proposed by ref. \cite{cd11}.

All told, our model space for treating the $11/2^-$ isomers consists of the following five configurations: $\tilde{S}=\phi_{\nu}\otimes S_{\pi}$, $\tilde{D}=\phi_{\nu}\otimes D_{\pi}$, $\tilde{G}=\phi_{\nu}\otimes G_{\pi}$, $\tilde{I}=\phi_{\nu}\otimes I_{\pi}$, and $\tilde{K}=\phi_{\nu}\otimes K_{\pi}$.

The $Q$ and $\mu$ matrix elements for the neutron configurations can be calculated according to the seniority property of the $\phi_{\nu}$ configuration, as described in ref. \cite{cd11}. For the neutron $Q$ matrix element,
\begin{equation}
\langle \phi_{\nu}~|~\hat{Q}~|~\phi_{\nu}\rangle=\frac{6-n}{5}~Q_{sp}(h_{11/2}),
\end{equation}
where $Q_{sp}(h_{11/2})=-0.055 r^2_0 \times e_{\nu}$ is the single-particle $Q$ value of the $h_{11/2}$ orbit, as estimated in ref. \cite{cd11}, $r^2_0=\frac{\hbar}{M\omega}=1.012A^{1/3}$, $e_{\nu}$ is the neutron effective charge, and $n=1+\frac{5}{9}(A-111)$ is the occupation number of the $h_{11/2}$ orbit as assumed by ref. \cite{cd11}. For the neutron $\mu$ matrix element,
\begin{equation}
\langle \phi_{\nu}~|~\hat{\mu}~|~\phi_{\nu}\rangle=\mu_{sp}(h_{11/2})=-1.339\mu_N,
\end{equation}
where $\mu_{sp}(h_{11/2})$ is the Schmidt $\mu$ value of the $h_{11/2}$ orbit, with the neutron spin Lande factor $g_{s\nu}=-3.827\times 0.7$.

The proton $Q$ and $\mu$ matrix elements of the $(g_{9/2})^{-2}$ configurations can be calculated with the formalism of ref. \cite{npa-for}. Additionally, appropriate proton effective charge $e_{\pi}$ and Lande factors are essential to reproduce the electromagnetic moments of the $I^{\pi}=11/2^-$ isomers and we now describe how we choose them. The $e_{\pi}$ value is determined by the experimental $2^+_1$ $Q$ value of $^{112}$Cd \cite{ensdf}. Due to the $N=64$ subshell closure \cite{N64-0,N64-1,N64-2,N64-3,N64-4,N64-5}, the $2^+_1$ state of ``semi-magic" $^{112}$Cd can be represented by a relatively pure $D_{\pi}$ configuration. Thus, the $Q$ value of the $2^+_1$ state corresponds to the $\langle D_{\pi}||\hat{Q}||D_{\pi}\rangle$ matrix element.
To match the calculated matrix element to the experimental $Q$ value, $e_{\pi} = -2.844$e is required, and this is adopted for all the other calculated $Q$ matrix elements in this work.
On the other hand, the experimental B(E2, $2^+_1\rightarrow 0^+_1$) and B(E2, $4^+_1\rightarrow 2^+_1$) values of $^{112}$Cd are also available for an estimate of the $\langle S_{\pi}||\hat{Q}||D_{\pi}\rangle$ and $\langle D_{\pi}||\hat{Q}||G_{\pi}\rangle$ matrix elements beyond the $N=64$ subshell closure. However, due to the strong collectivity of the Cd isotopes \cite{oct}, the adopted $e_{\pi}$ value can not adequately reproduce these two B(E2) values with simple $(g_{9/2})^{-2}$ configurations. To account for such collectivity, we directly extract magnitudes of the $\langle S_{\pi}||\hat{Q}||D_{\pi}\rangle$ and $\langle D_{\pi}||\hat{Q}||G_{\pi}\rangle$ matrix elements from the corresponding experimental B(E2) values of $^{112}$Cd, and adopt them in the calculations to follow. The proton Lande factors are chosen as usual to be $g_{l\pi}=1$ and $g_{s\pi}=5.586\times 0.7$, and are then used in the calculation of the proton $\mu$ matrix elements. All the adopted proton $Q$ and $\mu$ matrix elements based on these considerations are listed in Table \ref{pqm}.

\begin{table}
\caption{Adopted proton electromagnetic reduced matrix elements used in this work (see text for a detailed description of the calculation).}\label{pqm}
\begin{tabular}{lccccccccccccccccccccccccccccccc}
\hline\hline
$\langle D_{\pi}||\hat{Q}|| D_{\pi}\rangle$		&	 $-4.475 e\times r^2_0$	&	$\langle D_{\pi}||\hat{\mu}|| D_{\pi}\rangle$		&	 $+3.698 \mu_N$ 	\\
$\langle G_{\pi}||\hat{Q}|| G_{\pi}\rangle$		&	 $-2.134 e\times r^2_0$	&	$\langle G_{\pi}||\hat{\mu}|| G_{\pi}\rangle$		&	 $+6.751 \mu_N$ 	\\
$\langle I_{\pi~}||\hat{Q}|| I_{\pi~}\rangle$		&	 $+1.207 e\times r^2_0$ 	&	$\langle I_{\pi~}||\hat{\mu}|| I_{\pi~}\rangle$		&	 $+9.783 \mu_N$ 	\\
$\langle K_{\pi}||\hat{Q}|| K_{\pi}\rangle$		&	 $+5.711 e\times r^2_0$	&	$\langle K_{\pi}||\hat{\mu}|| K_{\pi}\rangle$		&	 $+12.809 \mu_N$ 	\\
$\langle S_{\pi}||\hat{Q}|| D_{\pi}\rangle$		&	 $-14.265 e\times r^2_0$	&		&		\\
$\langle D_{\pi}||\hat{Q}|| G_{\pi}\rangle$		&	 $-12.164 e\times r^2_0$	&		&		\\
\hline\hline
\end{tabular}
\end{table}

Due to the predominance of $S_{\pi}$ and $D_{\pi}$ configurations in the Cd low-lying states \cite{cd-fu}, we first consider trial wave functions as $C_S\tilde{S} + C_D\tilde{D}$ for the $I^{\pi}=11/2^-$ isomer of the odd-mass Cd isotopes, where $C_S^2$ and $C_D^2$ are wave function amplitudes. These amplitudes are numerically determined by
\begin{eqnarray}
&&C_S^2+C_D^2=1 ~.\\ \nonumber
&&C_S^2\langle \tilde{S}|\hat{\mu}|\tilde{S}\rangle+C_D^2\langle \tilde{D}|\hat{\mu}|\tilde{D}\rangle=\mu_{\rm exp} ~, \nonumber
\end{eqnarray}
where $\mu_{\rm exp}$ corresponds to the experimental $\mu$ values of the $I^{\pi}=11/2^-$ isomers. In other words, $\mu_{\rm exp}$ is an experimental constraint to the trial wave functions. Calculated values for $C_S^2$ and $C_D^2$ along with $e_{\nu}$ are taken as input to the $Q$ calculation for the $11/2^-$ isomer, with both options for the $C_D$ sign relative to $C_S$ considered and with the eventual choice depending on which result is closer to experiment. We fit the calculated $Q$ results to the experimental values, with $e_{\nu}$ as a further fitting parameter, and with $\mu_{\rm exp}$ as a constraint. This is referred to as our ``$SD$ fit". We present the optimal $e_{\nu}$ and the root-mean-square deviation (RMSD) of the $SD$ fit in Table \ref{fit}. The resulting $e_{\nu} = 7.334$e is too large, and the 0.456{\it eb} RMSD is of the same order as the experimental $Q$ values, which is clearly not satisfactory. This suggests that additional proton configurations are required.

\begin{table}
\caption{Fitting results with $SD$, $SDG$, $SDI$ and $SDK$ trial wave functions. $a$ and $b$ are linear parameters of $C_D$ (see Eq. \ref{sdgik} for their definition).}\label{fit}
\begin{tabular}{lcccccccccccccccccccccccccccccc}
\hline\hline
	&	$e_{\nu}$ (e)	&	$a$	&	$b$	&	RMSD (eb)	\\
	\hline
$SD$	&	7.334$\pm$2.110	&		&		&	0.456	\\
$SDG$	&	1.496$\pm$0.221	&	$+$0.017$\pm$0.004	&	$-$0.213$\pm$0.035	&	0.009	\\
$SDI$	&	3.079$\pm$6.479	&	$-$0.008$\pm$0.119	&	$+$0.036$\pm$1.189	&	0.018	\\
$SDK$	&	3.073$\pm$10.44	&	$-$0.010$\pm$0.194	&	$+$0.047$\pm$1.943	&	0.018	\\
\hline\hline
\end{tabular}
\end{table}

To accommodate the need for further proton configurations, we consider three other trial wave-functions: $C_S\tilde{S} + C_D\tilde{D} + C_G\tilde{G}$, $C_S\tilde{S} + C_D\tilde{D} + C_I\tilde{I}$ and $C_S\tilde{S} + C_D\tilde{D} + C_K\tilde{K}$ (denoted by ``$SDG$, $SDI$ and $SDK$", respectively). Here too $C_G$, $C_I$ and $C_K$ are wave function amplitudes to be constrained by the $\mu_{\rm exp}$ values, as for the $SD$ fit. Since the $D_{\pi}$ configuration should be the most important first-order proton excitation in Cd isotopes, the linear variation of the $11/2^-$ $Q$ and $\mu$ values suggests a linear behavior of $C_D$ in these isomers. Thus, we parameterize $C_D$ linearly in all three trial wave functions, $SDG$, $SDI$ and $SDK$. Taking the $SDG$ wave-function as an example, its amplitudes follow

\begin{eqnarray}\label{sdgik}
&&C_S^2+C_D^2+C_G^2=1,\\ \nonumber
&&C_D=a\times (A-110)+b,\\ \nonumber
&&C_S^2\langle \tilde{S}|~\hat{\mu}~|\tilde{S}\rangle+C_D^2\langle \tilde{D}|~\hat{\mu}~|\tilde{D}\rangle+C_G^2\langle \tilde{G}|~\hat{\mu}~|\tilde{G}\rangle=\mu_{\rm exp},
\end{eqnarray}
where $a$ and $b$ are the parameters that govern the linear behavior of $C_D$. The corresponding amplitudes for the $SDI$ and $SDK$ wave functions have similar relations to Eq. \ref{sdgik}. For given $a$, $b$ and $e_{\nu}$ values, different phases of $C_G$ (or $C_I$, $C_K$) relative to $C_S$ provide different $Q$ values. One can fix the $C_S$ sign to be positive, and the $C_G$ (or $C_I$, $C_K$) sign is then chosen so as to obtain a $Q$ value closest to the experimental value. Conversely, a fitting process to experimental $Q$ values optimizes the $SDG$, $SDI$ and $SDK$ wave-function with best-fit $a$, $b$ and $e_{\nu}$. Corresponding final best-fit values and the RMSD are also listed in Table \ref{fit}. All three fits achieve reasonable levels of agreement, with RMSD$\sim$ 0.01{\it eb}. Therefore, in what follows we limit ourselves to these trial wave functions.

When comparing the various fits, we conclude that the $SDG$ fit seems to be the best for the following three reasons:
\begin{itemize}
\item
The $SDG$ fit provides the smallest RMSD.
\item
The fitting errors of $a$, $b$ and $e_{\nu}$ in the $SDG$ analysis are two orders smaller than for either the $SDI$ or $SDK$ fits.
\item
The $SDI$ and $SDK$ fits also yield an unusually large $e_{\nu}\sim3$e value, as did the $SD$ fit discussed earlier.
\end{itemize}
Therefore, we only consider the $SDG$ wave-function with best-fit $a$, $b$ and $e_{\nu}$ values in the analysis to follow.

\section{analysis}
For the $SDG$ wave functions, the choice of the $C_G$ phase relative to that of $C_S$ can lead to different $Q$ values. However, the $SDI$ and $SDK$ fits have no such property. The enhanced flexibility associated with this double-valued behavior provides the $SDG$ wave function a greater opportunity to achieve a quality fit, suggesting why it is the best of the three. We classify the $SDG$ wave functions according to the relative phase of $C_G$ with respect to $C_D$ and $C_S$, with  $SDG+$ having $C_G>0$ and $SDG-$ having $C_G<0$.

In fig. \ref{q}, the calculated $Q$ values associated with the $SDG+$ and $SDG-$ wave-functions are compared with the experimental $Q$ values. The $SDG+$ results are compatible with the best-fit linear relation of the experimental $Q$ values, and provide a better description for the $N<70$ $Q$ values. In contrast, the $SDG-$ results exhibit some curvature beyond $N=70$, as in the experimental data. From these results, we conclude that there is a $C_G$ phase change across $N=70$. This provides a plausible explanation for the change in $Q$ linearity on the two sides of $N=70$, as reflected in fig. \ref{q}.

The precise difference between the $SDG+$ and $SDG-$ $Q$ values is observable but not very strong. It is desirable therefore to use other electromagnetic features to isolate the $C_G$ phase in the various isotopes. B(E2) values are normally more sensitive to the wave function than  $Q$ values. Thus, the systematic strong B(E2, $7/2^-_1\rightarrow 11/2^-_1$) values known for $^{113-119}$Cd \cite{ensdf} may provide additional confirmation of the $C_G$ phase change suggested by the $Q$ values above. We will thus calculate these B(E2) values, with both $SDG+$ and $SDG-$ wave-functions, and carry out a comparison between the experimental and calculated results.

As necessary input to the B(E2) calculation, we must first identify the main configuration of the initial $7/2^-_1$ state.  The strong E2 transition  $7/2^-_1\rightarrow 11/2^-_1$ empirically implies a similarity between the $11/2^-$ isomeric state and the $7/2^-_1$ state. This suggests treating the $7/2^-_1$ state as a proton-neutron recoupling of the $11/2^-$ isomer, namely as $\phi_{\nu}\otimes D_{\pi}$ or $\phi_{\nu}\otimes G_{\pi}$. As a first guess, the former wave function would seem preferable, since the $D_{\pi}$ configuration is energetically lower than the $G_{\pi}$ configuration and thus should dominate in the lowest state. Nevertheless, we also carried out test B(E2) calculations based on an initial $\phi_{\nu}\otimes G_{\pi}$ state, finding that it yields an $\sim$1 W.U. B(E2) value much smaller than found experimentally. Thus, we will assume that the primary configuration for the $7/2^-_1$ state is $\phi_{\nu}\otimes D_{\pi}$.

\begin{figure}
\includegraphics[angle=0,width=0.45\textwidth]{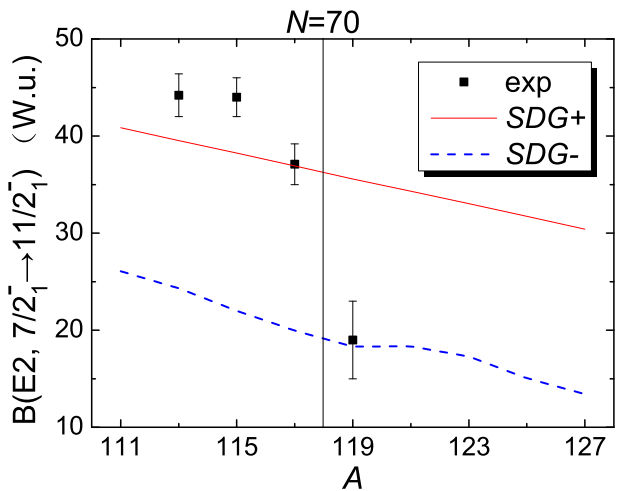}
\caption{(Color online) Calculated and experimental B(E2, $7/2^-_1\rightarrow 11/2^-_1$) values. Calculated results based on both the $SDG+$ and $SDG-$ wave functions for the $11/2^-$ isomers are presented. The experimental data (exp) is from ref. \cite{ensdf}. The sudden $N=70$ phase change is highlighted.}\label{be2}
\end{figure}

\begin{figure}
\includegraphics[angle=0,width=0.45\textwidth]{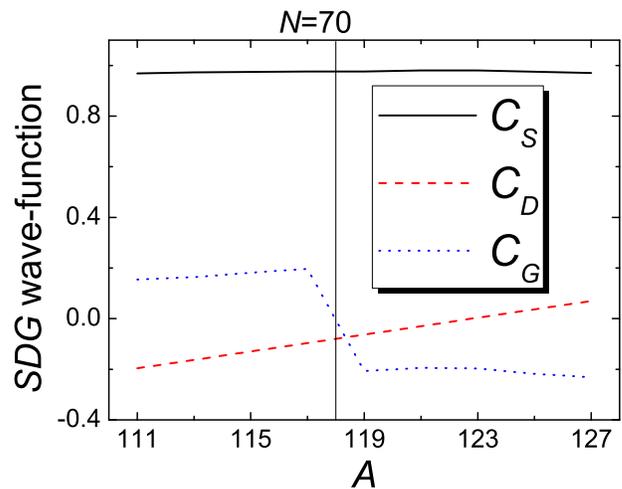}
\caption{(Color online) Wave functions from the final $SDG$ fit to the quadrupole moments and BE2 values. The $N=70$ phase change in $C_G$ is highlighted.}\label{wav}
\end{figure}

The calculated B(E2, $7/2^-_1\rightarrow 11/2^-_1$) values that emerge from both the $SDG+$ and $SDG-$ wave functions are compared with the available experimental data in fig. \ref{be2}. This comparison confirms a sudden phase change of $C_G$ at $N=70$: the $N<70$ B(E2) values clearly favor the $SDG+$ results ($C_G>0$), whereas for $N>70$, only the $SDG-$ calculation can fit the experimental B(E2) data. Using the different phases before and after $N=70$, we summarize our final $SDG$ wave functions from the $Q$-fitting procedure in fig. \ref{wav}.

What we see from this figure is that the $C_S$ value is always close to 1 for all the Cd isotopes. Thus, the $11/2^-$ isomer is constructed with a dominant $h_{11/2}$ single-particle configuration, with perturbations arising from $D_{\pi}$ and $G_{\pi}$ excitations. The second feature to note is that the $D_{\pi}$ contributions vary smoothly, as assumed earlier when we treated it as a linear function of neutron number (see Eq. \ref{sdgik}). On the other hand, we note that it passes through zero smoothly very near $N=70$, where $C_G$ undergoes its very rapid change of phase.

According to perturbation theory, changes in the $C_D$ and $C_G$ phases reflect sign changes of the associated neutron-proton interaction matrix elements. Due to the simplicity of the assumed proton configurations, such interactions must correspond to scalar products of multipole operators in the quadrupole and hexadecapole channels, respectively. The proton $(g_{9/2})^2$ configuration does not evolve with increasing neutron number, and thus the matrix elements of the proton multipole operators in such product states remain constant. Thus, it is the matrix element of the neutron operators that must change sign near $N=70$ to produce the observed behavior. As is well known \cite{Lawson}, a sign change in the matrix elements of an even multipole operator arises at the middle of a shell or isolated and degenerate subshell due to a transition from particle to hole behavior.

In ref. \cite{cd11}, the four neutron  orbits $s_{1/2}$, $d_{3/2}$, $d_{5/2}$ and $h_{11/2}$ of relevance to the Cd isotopes  were assumed to be nearly degenerate and reasonably well decoupled from the $g_{7/2}$ orbit. Under this assumption, the region from $N=58$ and $N=82$ provides a fairly well isolated subshell, with the middle of that subshell occurring at $N=70$ so that the particle-hole transition property for even multipole operators produces a phase change for that number of neutrons. It should be emphasized that this occurs for matrix elements involving any of the degenerate orbits, but most importantly for the $h_{11/2}$ orbital of relevance to this work. This would lead to a phase change in both the quadrupole and hexadecapole channels, but since $C_D$ changes sign so smoothly near $N=70$ (see fig. \ref{wav}) it shows up  most dramatically in the hexadecapole channel. Finally, we should note that not only does this mechanism seem to provide a natural explanation for the results presented in this work, but also provides added support for the shell structure assumed in ref. \cite{cd11}.

\section{summary}\label{sum}

To summarize, we have extracted shell-model wave functions of the $11/2^-$ isomers in the odd-mass $^{111-127}$Cd according to their electromagnetic features, including electromagnetic moments and B(E2) values. A phase change in the hexadecapole component of the extracted wave functions explains the change in linear behavior of the $Q$ and $\mu$ values for $N<70$ and $N>70$ as well as the B(E2, $7/2^-_1 \rightarrow 11/2^-_1$) behavior. Perturbative arguments suggest that the phase change near $N=70$ is related to the half-filling of the an isolated subshell with nearly degenerate $s_{1/2}$, $d_{3/2}$, $d_{5/2}$ and $h_{11/2}$ orbits, as proposed in ref. \cite{cd11}.

\acknowledgements
The authors gratefully acknowledge fruitful discussions with Dr. S. Q. Zhang, and constructive suggestions from an anonymous referee. This work was supported by the National Natural Science Foundation of China under contracts \# 11305151, 11305101, 11247241, and in part by the US National Science Foundation under grant \# PHY-0854873 0553127. One of the authors (J. H.) thanks the Shanghai Key Laboratory of Particle Physics and Cosmology for financial support (grant \# 11DZ2260700).

\end{document}